\documentclass{article}

\usepackage[final, nonatbib]{neurips_2021}

\usepackage[utf8]{inputenc} 
\usepackage[T1]{fontenc}    
\usepackage{hyperref}       
\usepackage{url}            
\usepackage{booktabs}       
\usepackage{amsfonts}       
\usepackage{nicefrac}       
\usepackage{microtype}      
\usepackage{xcolor}         
\usepackage{cite}
\usepackage{subcaption} 
\usepackage{graphicx}
\usepackage{multirow}
\usepackage{amsmath}
\usepackage{array}
\newcolumntype{P}[1]{>{\centering\arraybackslash}p{#1}}
\newcolumntype{M}[1]{>{\centering\arraybackslash}m{#1}}
\newcommand{\ie}{\textit{i.e.},\ }
\newcommand{\eg}{\textit{e.g.},\ }

\title{Reinforced Workload Distribution Fairness}

\author{%
  Zhiyuan Yao\\
  Department of Computer Science\\
  \'Ecole Polytechnique\\
  91120 Palaiseau, France \\
  \texttt{zhiyuan.yao@polytechnique.edu} \\
   \And
   Zihan Ding \\
   Department of Electrical and Computer Engineering\\
   Princeton University\\
   08544 New Jersey, U.S.A. \\
   \texttt{zhding96@gmail.com} \\
   \And
   Thomas Clausen\\
    Department of Computer Science\\
      \'Ecole Polytechnique\\
      91120 Palaiseau, France \\
  \texttt{thomas.clausen@polytechnique.edu} \\
}

\begin{document}

\maketitle

\begin{abstract}
  Network load balancers are central components in data centers, that distributes workloads across multiple servers and thereby contribute to offering scalable services.
  However, when load balancers operate in dynamic environments with limited monitoring of application server loads, they rely on heuristic algorithms that require manual configurations for fairness and performance.
  To alleviate that, this paper proposes a distributed asynchronous reinforcement learning mechanism to -- with no active load balancer state monitoring and limited network observations -- improve the fairness of the workload distribution achieved by a load balancer.
  The performance of proposed mechanism is evaluated and compared with state-of-the-art load balancing algorithms in a simulator, under configurations with progressively increasing complexities.
  Preliminary results show promise in RL-based load balancing algorithms, and identify additional challenges and future research directions, including reward function design and model scalability.
\end{abstract}

\section{Introduction}
\label{sec:intro}
In data centers and distributed systems, application servers are deployed on infrastructures with multiple servers, each with multiple processors to provide scalable services~\cite{dragoni2017microservices}.
To optimize workload distribution and reduce additional queuing delay, load balancers (LBs) play a significant role in such systems, yet rely on heuristics for decisions on where to place each workload~\cite{lvs, maglev, 6lb, incab2018}.
Reinforcement learning (RL) approaches~\cite{auto2018sigcomm, decima2018, drl-udn-2019, sivakumar2019mvfst} have shown performance gains in various system and networking problems.
They help avoid error-prone manual configurations.
This paper therefore studies the potential performance improvement when applying RL techniques on network load balancing problems in distributed systems.

Applications of RL techniques on network load balancing problems are challenging for several reasons.
First, unlike traditional workload distribution problem~\cite{auto2018sigcomm, decima2018}, network LBs have limited observations on task size and actual server load states.
Being unaware of task size, application servers can be overloaded by collided ``elephant tasks'' and thus yield degraded quality of service.
As network LBs operate on the Transport Layer, obtaining task size information before allocating workloads requires application-specific -- and thus by definition non-universal -- LB implementations~\cite{haproxy2012, nginx2008}.
With services running on heterogeneous hardware or elastic data centers~\cite{kumar2020fast} where server capacities vary, ``fairly'' (uniformly) distributing workloads without considering actual server load-states may overload servers with limited resources and underload powerful servers.
On the other hand, probing actual server load states requires deploying agents on servers and incurs additional management traffic~\cite{6lb, incab2018}.
Second, the interactive training procedure of RL models produces periodically and asynchronously updated actions in a slow pace (\eg every hundreds of milliseconds~\cite{sivakumar2019mvfst}) in the control plane, while system dynamics can change rapidly (\eg sub-millisecond in modern networked systems~\cite{guo2015pingmesh}) in the data plane.

This paper proposes an RL-based network load balancing mechanism (RLB-SAC) that is able to exploit asynchronous actions based only on local observations and inference.
Performance is evaluated on an event-based simulator and compared with benchmark load balancing mechanisms.

\section{RL for Network Load Balancing}

Network load balancing can be defined as allocating a Poisson sequence of tasks with different workloads $\{w_k\}\in\mathcal{W}$\footnote{The unit of workload can be, \eg amount of time to process.} on a set of $n$ servers, to achieve the maximal exploitation of the computational capacity of the servers.
The task workload $w^j_k(t)$ assigned on the $j$-th server at time $t$ usually follows an exponential distribution in practical experiments~\cite{facebook-dc-traffic}. 
The load balancing method is a function $\pi\in\Pi$: $\mathcal{W}\rightarrow [n]$. 
The processing speed for each server is $s_j, j\in[n]$, \ie the amount of workloads that can be processed per unit of time. 
The load on the $j$-th server ($j\in[n]$) during a time interval $t\in[t_0, t_n]$ is thus defined as: 
\begin{align}
    l_j=\frac{\sum_{t\in[t_0, t_n)}w^j_k(t)}{s_j},
\end{align}
which represents the expected time to finish processing all the workloads on the $j$-th server.

The objective of load balancing can be defined as finding the optimal policy:
\begin{align}
    \pi^* = \min_{\pi\in\Pi} \max_j l_j
\end{align}

This problem is multi-commodity flow problems and is NP-hard, which makes it hard to solve with trivial algorithmic solution within micro-second level~\cite{sen2013scalable}.

\subsection{RL Methodology}
\label{sec:rl}
\begin{figure}[t]
	\centering
	\centerline{\includegraphics[width=\columnwidth]{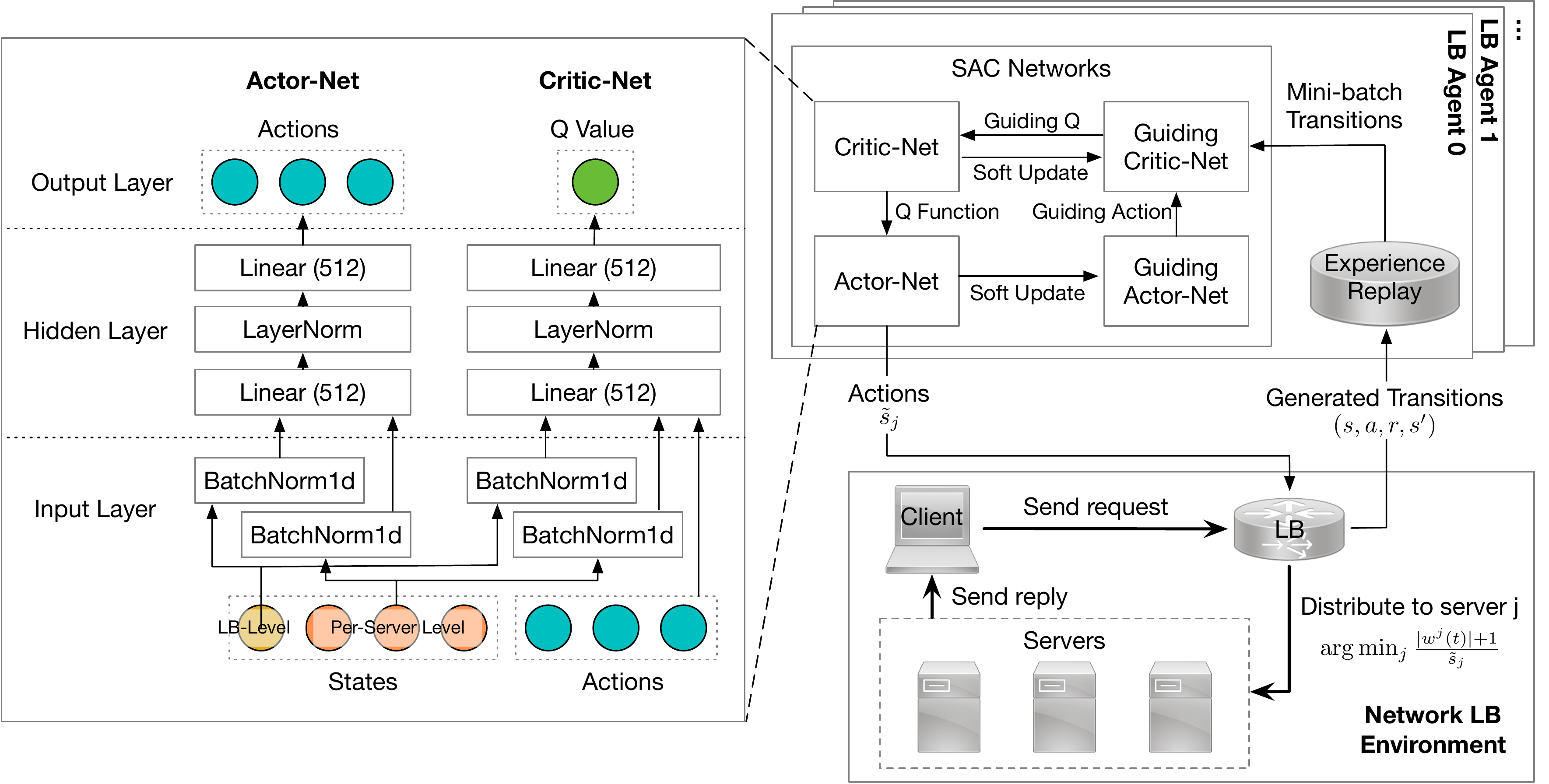}}
	\caption{Overview of the proposed RL framework for network LB. A distributed learning framework with multiple LB agents is implemented to interact with the simulated LB and allocate tasks on different servers according to a server assignment function. Each LB agent contains a replay buffer and a SAC model with two pairs of actor and critic networks. Network architectures are shown on the left block. Linear $(n)$ indicates the number of hidden units is $n$.}
	\label{fig:design}
	\vskip -.1in
\end{figure}

To apply standard RL methods, the network load balancing problem can be formulated as a Markov Decision Process (MDP), which can be represented as $(\mathcal{S}, \mathcal{A}, R, \mathcal{T}, \gamma)$. 
$\mathcal{S}$ and $\mathcal{A}$ are the state space and the action space, and $R$ is a reward function $R(s,a)$: $\mathcal{S}\times \mathcal{A}\rightarrow \mathbb{R}$ for current state $s\in\mathcal{S}$ and action $a\in\mathcal{A}$. 
The state-transition probability from current state and action to a next state $s^\prime\in\mathcal{S}$ is defined by $\mathcal{T}(s^\prime|s,a)$: $\mathcal{S}\times\mathcal{A}\times\mathcal{S}\rightarrow \mathbb{R}_+$. $\gamma\in(0,1)$ is a reward discount factor. The goal of an RL algorithm is optimizing the policy to maximize the agent's expected cumulative rewards: $\mathbb{E}[\sum_t \gamma^t r_t]$. 

Soft Actor-Critic (SAC)~\cite{haarnoja2018soft} follows the maximum entropy reinforcement learning framework, which optimizes the objective $\mathbb{E}[\sum_t \gamma^t r_t+\alpha \mathcal{H(\pi_\theta)}]$ to encourage the exploration $\mathcal{H}(\cdot)$ of the policy $\pi_\theta$ during the learning process. More concretely, the critic $Q$ network is updated with the gradient $\nabla_\phi\mathbb{E}_{s,a}[Q_\phi(s,a)-[R(s,a)+\gamma \mathbb{E}_{s^\prime}[V_{\tilde{\phi}}(s^\prime)]]]$, where $V_{\tilde{\phi}}(s^\prime)=\mathbb{E}_{a^\prime}[Q_{\tilde{\phi}}(s^\prime, a^\prime)-\alpha\log\pi_\theta(a^\prime|s^\prime)]$ and $Q_{\tilde{\phi}}$ is the target $Q$ network; the actor policy $\pi_\theta$ is updated with the gradient $\nabla_\theta\mathbb{E}_s[\mathbb{E}_{a\sim\pi_\theta}[\alpha\log\pi_\theta(a|s)-Q_\phi(s,a)]]$.

Other key elements of RL methods involve the observation, action and reward function, which are detailed as following:

\textbf{Observation.} Based on the limited information that is visible to network LBs, \textit{i.e.} task ID $k$, arrival and finish time of each task, the number of ongoing tasks $|w^j(t)|$ on the $j$-th server can be computed and the following $3$ observations are sampled, \ie task inter-arrival time, task duration, and task completion time (TCT).
Each sampled time-related feature channel is reduced to 5 scalars, \eg average, 90th-percentile, standard deviation, discounted average and weighted discounted average\footnote{Discounted average weights are computed as $0.9^{t^\prime-t}$, where $t$ is the sample timestamp and $t^\prime$ is the moment of calculating the reduced scalar.}.
Task inter-arrival time is computed at LB level, while task duration and TCT are gathered at per-server level.

\textbf{Action.} To bridge the difference between slow-paced action updates and high-speed network packets arrivals, the RL agent assign the $j$-th server to newly arrived task using the ratio of two factors:
\begin{align}
\arg \min_{j\in[n]} \frac{|w^j(t)|+1}{\tilde{s}_j},
\label{eq:action}
\end{align}
where the number of on-going tasks $|w^j(t)|$ helps track dynamic system server occupation and $\tilde{s}_j$ is the periodically updated RL-inferred server processing speed given observations, as the direct outputs of the LB agents. Eq.~\eqref{eq:action} is the server assignment function.

\textbf{Reward.} The reward is chosen as $1-F(\mathbf{\tilde{w}})$, where $\mathbf{\tilde{w}}$ is a list of discounted average of actual TCT on each server, and $F$ is a fairness measurement.
In this paper, $3$ fairness indices are studied, namely, Jain's, G's, and Bossaer's fairness index, which have the following forms respectively, $F_J(\mathbf{\tilde{w}}) = \frac{\overline{\mathbf{\tilde{w}}}^2}{\overline{\mathbf{\tilde{w}}^2}}$, $F_G = \prod_{j \in [n]}\sin(\frac{\pi\tilde{w}^j}{2\max(\mathbf{\tilde{w}})})$, $F_B = \prod_{j \in [n]}\frac{\tilde{w}^j}{\max(\mathbf{\tilde{w}})}$.

\textbf{Model.} The architecture of the proposed RL framework is depicted in figure~\ref{fig:design}. The SAC model within each LB agent contains a pair of actor and critic networks, as well as a pair of guiding actor and guiding critic networks, with the same network architectures as the former but updated in a delayed and soft manner. As shown in the left block of figure~\ref{fig:design}, the actor and critic networks take the batch-normalized features from LB-level states and per-server-level states and use the same Linear-LayerNorm-Linear layers to process the data independently. The critic additionally takes the actions as inputs to generate the estimated $Q$ values.

\section{Implementation}

In order to compare and contrast performance of different load balancing methods in various scenarios, in particular those that are difficult to evaluate in testbeds, such as large-scale data center networks, an event-driven simulator is implemented.

\subsection{Server Processing Model}
\label{sec:implement-simulator-server}

Realistic network applications feature blocked processor sharing model~\cite{hadoop2011apache, spark2018apache}, in which the instantaneous processing speed $s_j(t)$ at time $t$ on the $j$-th server is:
\begin{equation}
	 s_j(t)=
	\begin{cases}
	1       & \quad |w^{j}(t)| \leq p_{j},\\
	\frac{p_{j}}{\min \left(\hat{p}_{j}, |w^{j}(t)|\right)}  & \quad |w^{j}(t)| > {p}_{j},
	\end{cases}
\end{equation}
where $|w^{j}(t)|$ denotes the number of on-going tasks, and $p_j$ denotes the number of processors on the $j$-th server.
At any given moment, the maximum number of tasks that can be processed is $\hat{p}_j$.
Tasks that arrive when $|w^{j}(t)| \geq \hat{p}_j$ will be blocked in a queue (similar to the backlog in \eg Apache), and will not be processed until there is an available slot in the CPU processing queue.

\subsection{Benchmark Load Balancing Methods}
\label{sec:simulator-methods}

To quantify the performance of the proposed LB method (RLB-SAC), $4$ baseline workload distribution algorithms are implemented.
Equal-cost multi-path (ECMP) randomly assigns servers to tasks with a server assignment function $\mathbb{P}(j) = \frac{1}{n}$, where $\mathbb{P}(j)$ denotes the probability of assigning the $j$-th server~\cite{faild2018}.
Weighted-cost multi-path (WCMP) assigns servers based on their weights derived, and has an assignment function as $\mathbb{P}(j) = \frac{p_j}{\sum p_j}$~\cite{maglev}.
Local shortest queue (LSQ) assigns the server with the shortest queue, \ie $\arg \min_{j\in[n]} |w^j(t)|$~\cite{twf2020}.
And finally, shortest expected delay (SED) assigns the server the shortest queue normalized by the number of processors, \ie $\arg \min_{j\in[n]} \frac{|w^j(t)|+1}{p_j}$~\cite{lvs}, and is expected to have the best performance among baseline LB methods.

\section{Experimental Evaluation}
\label{sec:experiment}

This section investigates the impacts of several key components for our proposed RL framework, including (i) the system dynamics properties like traffic rate and TCT (workload $w^j_k(t)$) distribution, (ii) the reward function in RL and (iii) the scalability of the system. Hyperparameters for training see Appendix.~\ref{app:hyperparameter}.

\begin{figure}[t]
	\centering
	\begin{subfigure}{.47\columnwidth}
		\centerline{\includegraphics[height=1.4in]{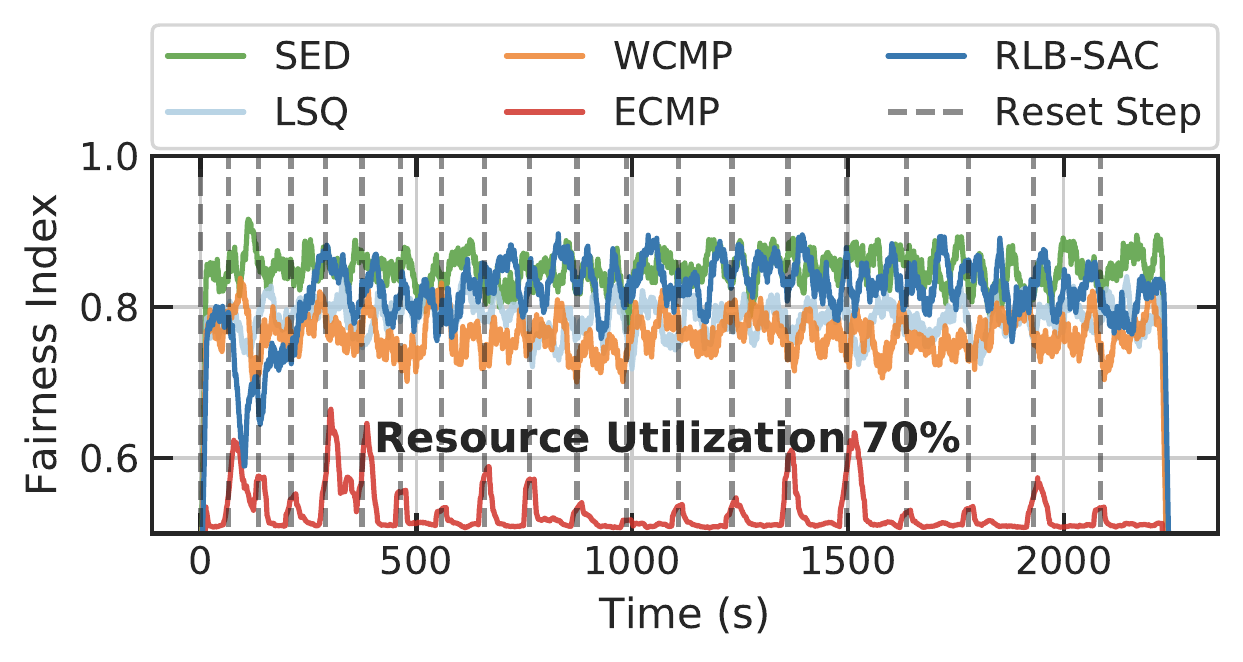}}
	    \vskip -.1in
		\caption{}
		\label{fig:timeline-rate-70}
	\end{subfigure}%
	\hspace{.1in}
	\begin{subfigure}{.47\columnwidth}
		\centerline{\includegraphics[height=1.4in]{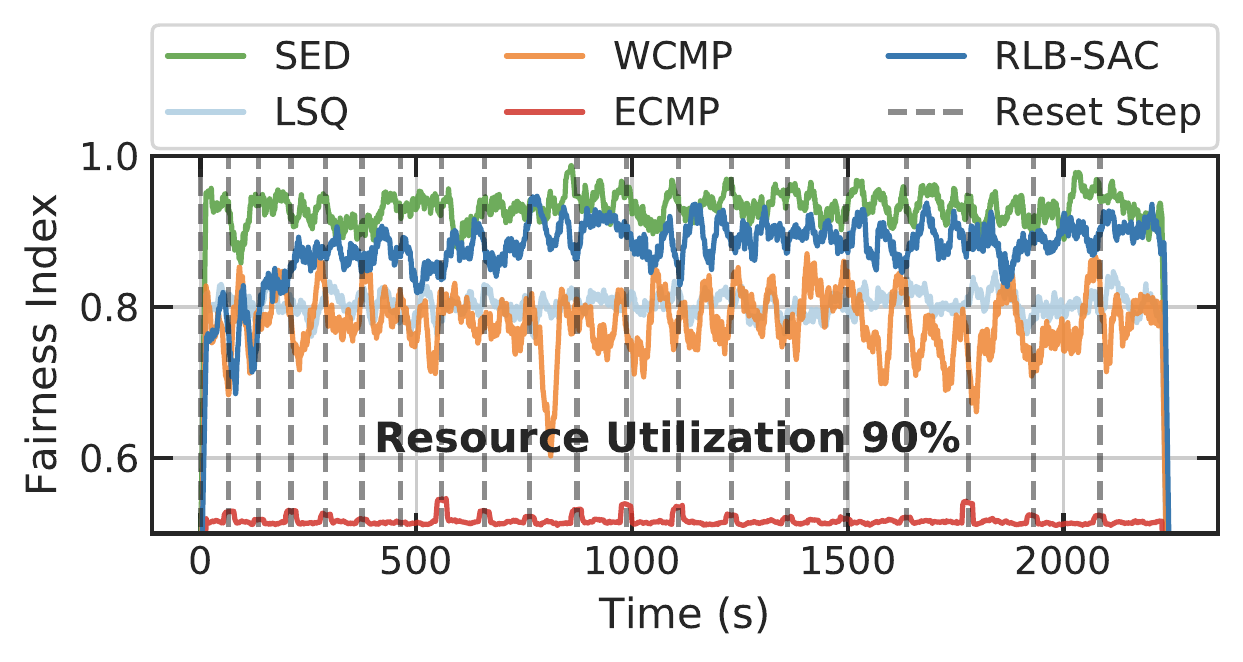}}
		\vskip -.1in
		\caption{}
		\label{fig:timeline-rate-90}
	\end{subfigure}%
	\vspace{-.02in}
	\begin{subfigure}{.47\columnwidth}
		\centerline{\includegraphics[height=1.4in]{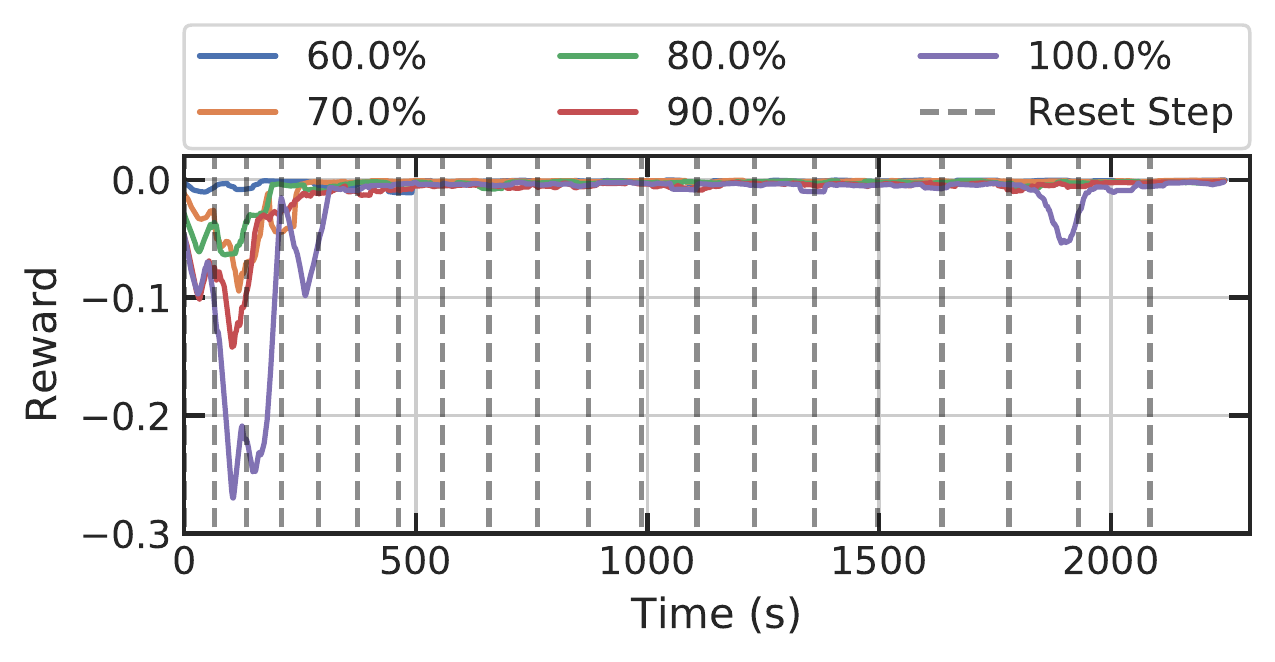}}
		\vskip -.1in
		\caption{}
		\label{fig:timeline-reward}
	\end{subfigure}
	\hspace{.1in}
	\begin{subfigure}{.47\columnwidth}
		\centerline{\includegraphics[height=1.4in]{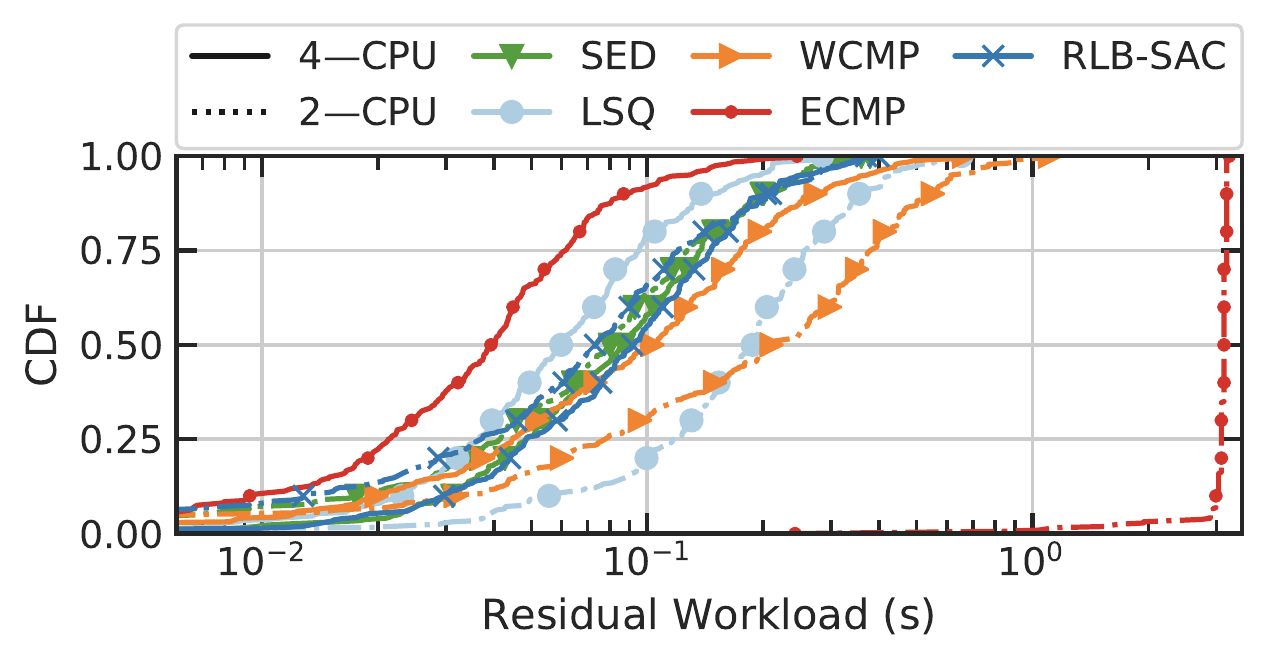}}
		\vskip -.1in
		\caption{}
		\label{fig:residual-workload-cdf}
	\end{subfigure}
	\vskip -.02in
	\caption{Comparison of load balancing performance (during training) under different traffic rates in a simplest setup where all tasks are the same.
  Figure(a) and (b) show Jain's fairness index of residual workloads $l_j$ between $2$ servers over time.
  Figure(c) shows the reward over $20$ episodes under different traffic rates.
  Figure(d) compares the CDF of residual workloads $l_j$ on $2$ servers in the last episode.}
	\label{fig:illustration}
	\vskip -.2in
\end{figure}

\begin{table}[t]
\centering
\caption{Comparison under different traffic rates ($60\%, 70\%, 80\%, 90\%, 100\%$) with the same TCT. ``FI'' and ``RW'' are ``Jain's fairness index'' and ``average residual workloads'' respectively in the last episode.}
\resizebox{\textwidth}{!}{ 
\begin{tabular}{c|c|c|c|c|c|c|c|c|c|c}
\toprule
\multirow{3}{*}{Method} & \multicolumn{10}{c}{Traffic Rate (\%)} \\
\cline{2-11}
 & \multicolumn{2}{c|}{60} & \multicolumn{2}{c|}{70} & \multicolumn{2}{c|}{80} & \multicolumn{2}{c|}{90}& \multicolumn{2}{c}{100} \\
\cline{2-11}
  & FI & RW & FI & RW & FI & RW & FI & RW & FI & RW\\
\hline
 ECMP & $0.646$ & $0.130$ & $0.512$ & $1.225$ & $0.512$ & $1.504$ & $0.515$ & $1.553$ & $0.517$ & $1.568$ \\
 WCMP & $0.753$ & $0.046$ & $0.765$ & $0.060$ & $0.788$ & $0.088$ & $0.787$ & $0.192$ & $0.835$ & $0.809$\\
 LSQ & $0.768$ & $0.046$ & $0.791$ & $0.062$ & $0.789$ & $0.090$ & $0.806$ & $0.137$ & $0.864$ & $0.607$ \\
 SED & $\mathbf{0.841}$ & $\mathbf{0.033}$ & $\mathbf{0.855}$ & $0.046$ & $\mathbf{0.902}$ & $0.067$ & $\mathbf{0.929}$ & $\mathbf{0.099}$ & $\mathbf{0.989}$ & $\mathbf{0.417}$\\
\textbf{RLB-SAC} & $0.782$ & $\mathbf{0.033}$ & $0.807$ & $\mathbf{0.040}$ & $0.874$ & $\mathbf{0.064}$ & $0.904$ & $0.100$ & $0.969$ & $0.520$ \\
\bottomrule
\end{tabular}
}
\vskip -.2in
\label{tab:compare_traffic_rate}
\end{table}

\textbf{Traffic Rate.} 
Following the approach in~\cite{fu2021use}, a simplest scenario, \ie all the tasks are identical (each has $100$ms TCT), is created to verify the learning ability of RLB-SAC.
Different traffic rates are applied on a system with $1$ LB node and $2$ servers, where one server has $4$ CPUs and the other has $2$ CPUs.
The traffic rates are normalized to $[0, 1]$ by the processing capacity of configured server clusters so that $100\%$ traffic rate creates a stable system state (equal average arrival and departure rate).
Jain's fairness index is used to compute reward and evaluate workload distribution performance.
The result of $20$-episode training in simulation\footnote{The first episode lasts $60$s and each following episode has $5$s incremental duration than the former one. The time interval between two consecutive steps is $0.5$s. The same configuration applies throughout the paper.} is depicted in figure~\ref{fig:illustration} and listed in table \ref{tab:compare_traffic_rate}.
After $5$ episodes, the proposed SAC-based LB (RLB-SAC) improves its performance in terms of workload distribution fairness.
RLB-SAC learns to behave similar to SED and assigns similar workloads $l_j$ across the two servers.
Though unlike SED, RLB-SAC requires no manual configuration.

\begin{figure}[t]
	\centering
	\begin{subfigure}{.47\columnwidth}
		\centerline{\includegraphics[height=1.4in]{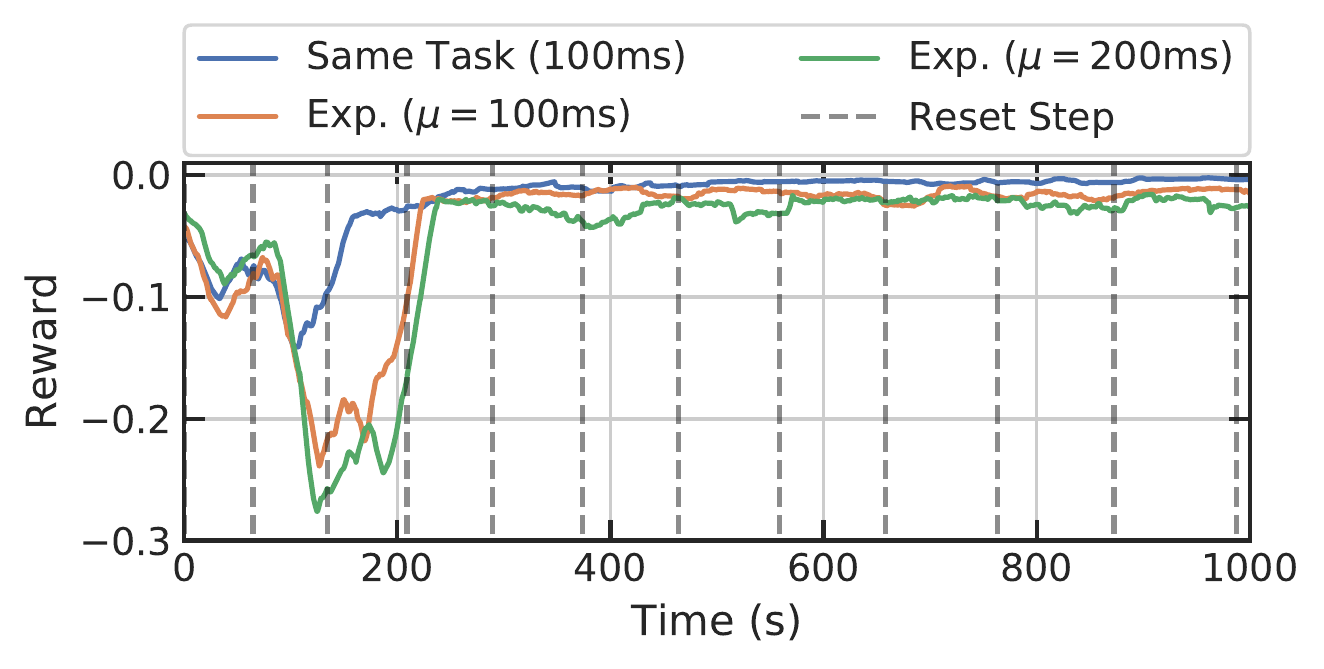}}
	    \vskip -.1in
		\caption{}
		\label{fig:exp-reward}
	\end{subfigure}%
	\hspace{.1in}
	\begin{subfigure}{.47\columnwidth}
		\centerline{\includegraphics[height=1.4in]{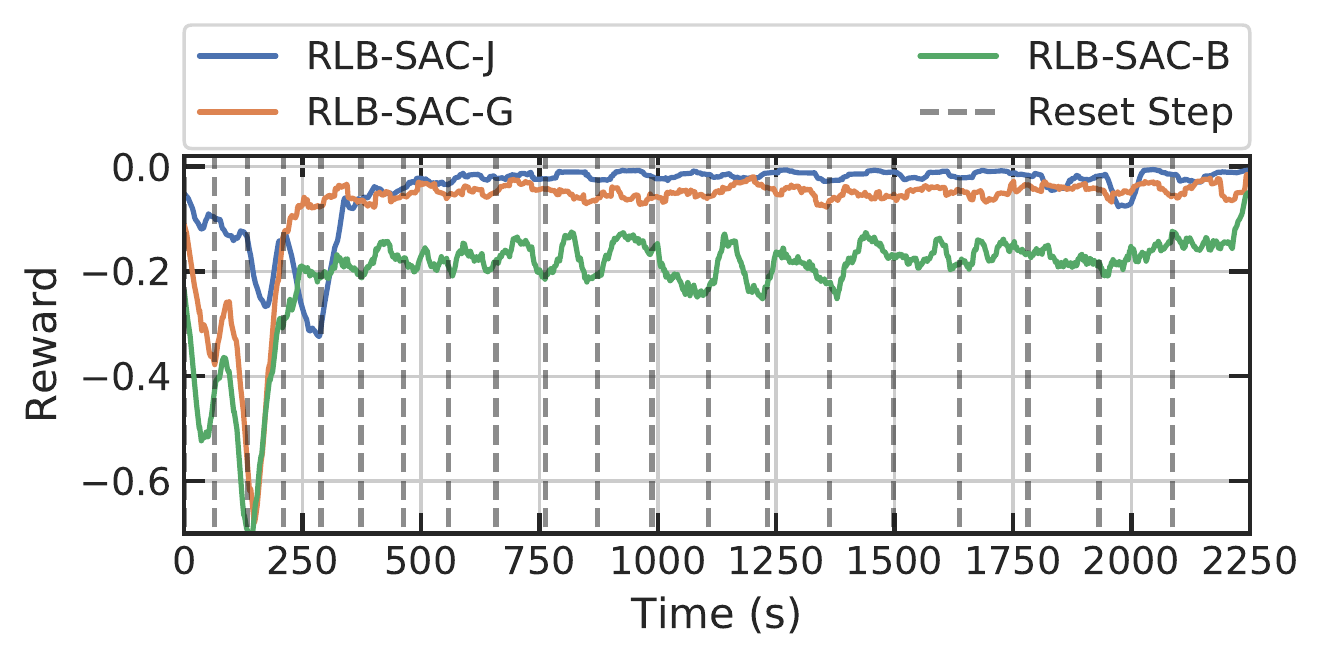}}
	    \vskip -.1in
		\caption{}
		\label{fig:exp-reward-option}
	\end{subfigure}%
	\vspace{-.02in}
	\begin{subfigure}{.47\columnwidth}
		\centerline{\includegraphics[height=1.4in]{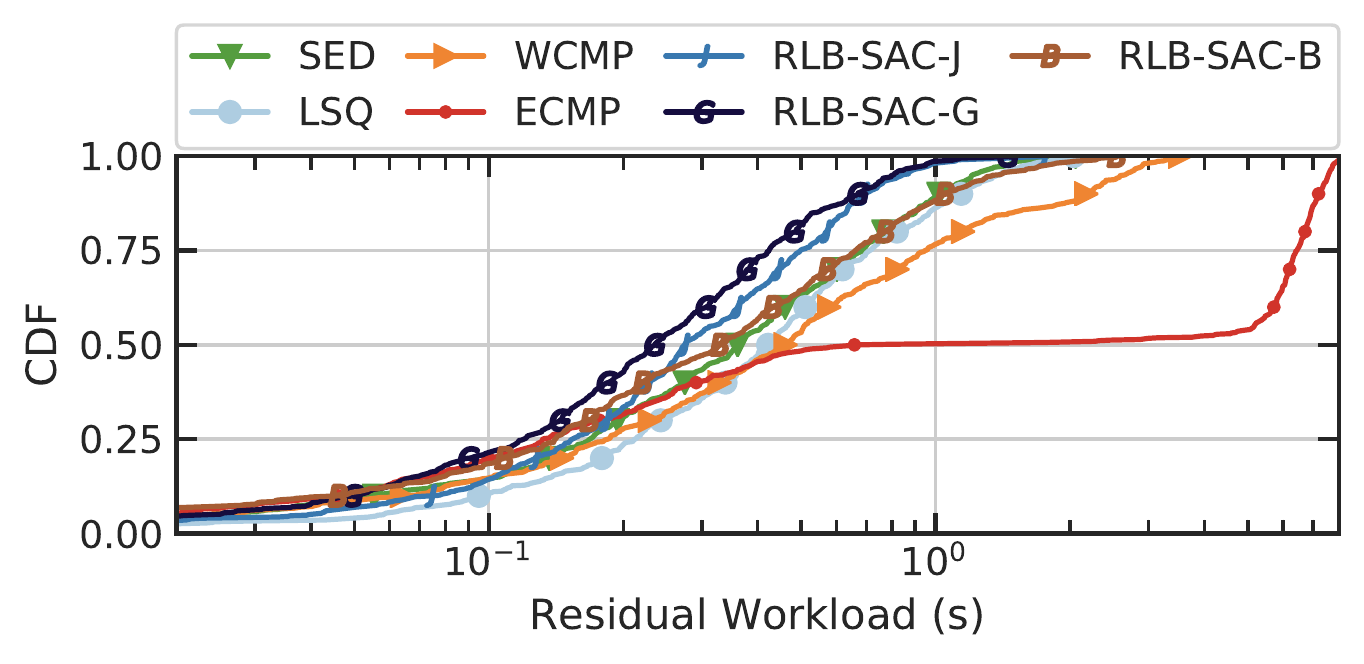}}
		\vskip -.1in
		\caption{}
		\label{fig:exp-option-rate-90}
	\end{subfigure}%
	\hspace{.1in}
	\begin{subfigure}{.47\columnwidth}
		\centerline{\includegraphics[height=1.4in]{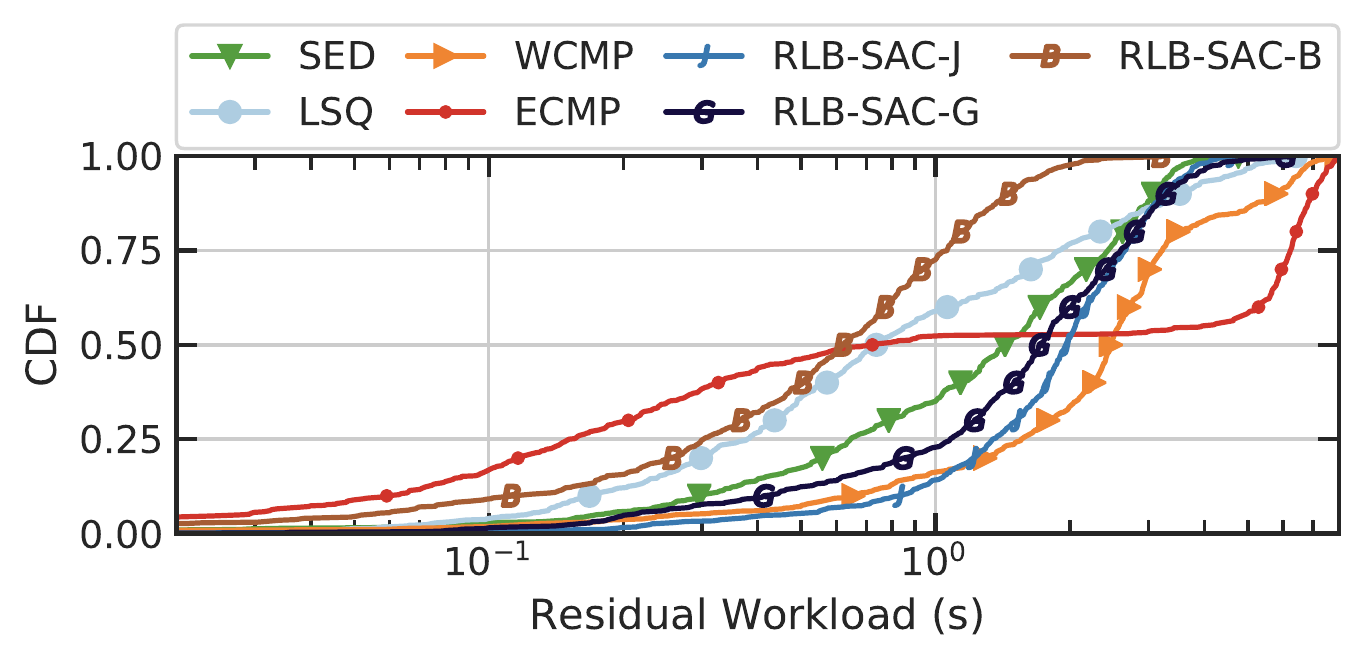}}
		\vskip -.1in
		\caption{}
		\label{fig:exp-option-rate-100}
	\end{subfigure}
	\vspace{-.02in}
	\caption{Figure(a) shows Jain's fairness index reward over $20$ episodes using different TCT distributions under $90\%$ traffic rate.
  Figure(b) shows different reward options under $100\%$ traffic rate and exponential TCT distributions ($\mu=200$ms).
  Figure(c) and (d) shows CDF of residual workloads using different LB mechanisms in the last episode, under $90\%$ and $100\%$ traffic of tasks with exponentially distributed TCT ($\mu=200$ms) respectively. All results are obtained during the training process.}
	\label{fig:exp-traffic}
	\vskip -.2in
\end{figure}

\begin{table}[t]
\caption{Comparison of the last episode performance under different traffic rates ($60\%, 70\%, 80\%, 90\%, 100\%$) with the exponential TCT distribution ($\mu=200$ms). ``Max'' and ``Avg'' stand for the maximum and average residual workload in the last episode.}
\centering
\resizebox{\textwidth}{!}{ 
\begin{tabular}{c|c|c|c|c|c|c|c|c|c|c}
\hline
\toprule
\multirow{3}{*}{Method} & \multicolumn{10}{c}{Traffic Rate (\%)} \\
\cline{2-11}
 & \multicolumn{2}{c|}{60} & \multicolumn{2}{c|}{70} & \multicolumn{2}{c|}{80} & \multicolumn{2}{c|}{90}& \multicolumn{2}{c}{100} \\
\cline{2-11}
  & Avg & Max & Avg & Max & Avg & Max & Avg & Max & Avg & Max \\
\hline
 SED & $\mathbf{0.127}$& $\mathbf{0.776}$& $0.183$& $1.055$ & $\mathbf{0.238}$ & $\mathbf{0.951}$& $0.463$& $2.058$& $1.571$& $4.762$\\
\hline
 LSQ & $0.173$ &$1.306$ & $0.225$ & $1.246$ & $0.304$ & $1.770$ & $0.522$& $2.038$& $1.352$& $6.408$\\
\hline
 WCMP & $0.159$&$1.013$ & $0.214$ & $1.360$ & $0.346$& $2.999$ & $0.739$& $3.531$& $2.705$& $7.896$\\
\hline
 ECMP & $0.662$& $3.085$ & $1.365$ & $5.565$ & $3.088$& $9.129$ & $3.233$& $8.158$& $3.032$& $8.048$\\
\hline
 \textbf{RLB-SAC-J} & $0.145$ & $0.952$ & $0.173$ & $1.188$ & $0.299$& $2.208$& $0.350$& $1.747$& $2.004$& $4.608$\\
\hline
 \textbf{RLB-SAC-G} & $0.162$ & $0.994$ & $\mathbf{0.162}$ & $\mathbf{0.766}$ & $0.245$& $1.573$& $\mathbf{0.301}$& $\mathbf{1.451}$& $1.846$& $6.072$\\
\hline
 \textbf{RLB-SAC-B} & $0.146$ & $1.072$ & $0.174$ & $1.082$ & $0.265$& $1.239$ & $0.464$& $2.538$& $\mathbf{0.722}$& $\mathbf{3.194}$\\
\bottomrule
\end{tabular}
}
\vskip -.2in
\label{tab:compare_reward_option}
\end{table}

\textbf{TCT Distribution.}
In a more realistic setup, tasks are different and follows long-tail distributions~\cite{facebook-dc-traffic}.
As is depicted in figure~\ref{fig:exp-reward}, the additional variance induced by exponential distribution TCT increases the difficulty of learning.
Since the $\mu$ of exponential distributions is equivalent to its standard deviation, increasing the average TCT (from $100$ms to $200$ms) also increases the standard deviation of TCT.
The additional variance in the TCT observation thus further reduces rewards obtained using RLB-SAC over time.

\textbf{Reward Function.} This section further compares different choices of the reward function in the RL procedure, including the Jain's (RLB-SAC-J), G's (RLB-SAC-G), and Bossaer's (RLB-SAC-B) fairness index as introduced in section~\ref{sec:rl}.
Compared against the baseline LB methods, depicted in figure~\ref{fig:exp-option-rate-90} and \ref{fig:exp-option-rate-100}, the SAC method with these three types of reward is tested under different traffic rates and exponential TCT distribution ($\mu=200$ms).
More complete results for the episodic maximum and average residual workload are listed in table~\ref{tab:compare_reward_option}.
It shows that, without any prior knowledge of server processing capacities, the SAC methods work significantly better than other methods for cases with high traffic rates ($90\%-100\%$), with close performance to the best SED method for relatively low traffic rates.
Moreover, RLB-SAC-G and RLB-SAC-B fairness index have superior performance over RLB-SAC-J for most cases.
This can be explained by the fact that Jain's fairness index provides values that are close to $1$ unless ``severe'' unfairness is present, which also makes the reward close to $0$.
Bossaer's fairness index yields lower values and is more sensitive to unfairness (as in figure~\ref{fig:exp-reward-option}), which makes it better when evaluating system state than G's fairness index especially when the traffic rate is higher (\eg $100\%$ as in figure~\ref{fig:exp-option-rate-100}).

\textbf{Scalability.}
To study the scalability of the RLB-SAC model, $4$ types of topology are applied where $1$ or $2$ LB agents distribute workloads across $4$ or $8$ servers, and half of all the servers have $4$ CPUs while the other half have $2$ CPUs.
As the action space grows, the task becomes more challenging and RLB-SAC fails to surpass SED after $40$ episodes of training.
Though RLB-SAC achieves the second place among all methods under moderate traffic rate ($70\%$).

\begin{table}[t]
\caption{Comparison of the last episode performance under $70\%$ traffic rate  with the exponential TCT distribution ($\mu=200$ms). ``Max'' and ``Avg'' stand for the maximum and average residual workload in the last episode.}
\footnotesize
\centering
\begin{tabular}{c|c|c|c|c|c|c|c|c}
\hline
\toprule
\multirow{3}{*}{Method} & \multicolumn{8}{c}{Network Topology} \\
\cline{2-9}
 & \multicolumn{2}{c|}{$1$LB-$4$S} & \multicolumn{2}{c|}{$2$LB-$4$S} & \multicolumn{2}{c|}{$1$LB-$8$S} & \multicolumn{2}{c}{$2$LB-$8$S} \\
\cline{2-9}
  & Avg & Max & Avg & Max & Avg & Max & Avg & Max \\
\hline
 SED & $\mathbf{0.146}$& $\mathbf{1.202}$ & $\mathbf{0.149}$ & $\mathbf{0.938}$& $\mathbf{0.140}$& $\mathbf{1.110}$& $\mathbf{0.141}$& $\mathbf{1.035}$ \\
\hline
 LSQ & $0.184$ &$1.661$ & $0.193$ & $1.302$ & $0.187$ & $1.518$ & $0.194$& $1.300$\\
\hline
 WCMP & $0.232$&$2.190$ & $0.223$& $2.803$ & $0.236$ & $2.822$ & $0.236$& $2.736$\\
\hline
 ECMP & $2.195$& $8.520$ & $2.256$& $8.604$ & $2.085$ & $9.483$ & $2.192$& $8.863$\\
\hline
 \textbf{RLB-SAC-J} & $0.177$ & $1.271$ & $0.186$& $1.385$ & $0.172$ & $1.741$ & $0.172$& $1.270$\\
\hline
 \textbf{RLB-SAC-G} & $0.167$ & $1.426$ & $0.188$& $1.725$ & $0.162$ & $1.334$ & $0.182$& $1.700$\\
\hline
 \textbf{RLB-SAC-B} & $0.167$ & $1.639$ & $0.175$& $1.695$ & $0.166$ & $1.653$ & $0.174$& $1.611$\\
\bottomrule
\end{tabular}
\label{tab:compare_reward_option}
\vskip -.2in
\end{table}

\begin{table}[t]
\caption{Comparison of the last episode performance under $90\%$ traffic rate  with the exponential TCT distribution ($\mu=200$ms). ``Max'' and ``Avg'' stand for the maximum and average residual workload in the last episode.}
\footnotesize
\centering
\begin{tabular}{c|c|c|c|c|c|c|c|c}
\hline
\toprule
\multirow{3}{*}{Method} & \multicolumn{8}{c}{Network Topology} \\
\cline{2-9}
    & \multicolumn{2}{c|}{$1$LB-$4$S} & \multicolumn{2}{c|}{$2$LB-$4$S} & \multicolumn{2}{c|}{$1$LB-$8$S} & \multicolumn{2}{c}{$2$LB-$8$S} \\
\cline{2-9}
    & Avg & Max & Avg & Max & Avg & Max & Avg & Max \\
\hline
    SED & $\mathbf{0.304}$ & $\mathbf{1.865}$ & $\mathbf{0.287}$ & $\mathbf{1.633}$ & $\mathbf{0.219}$ & $\mathbf{1.428}$ & $\mathbf{0.238}$& $\mathbf{1.467}$ \\
\hline
    LSQ & $0.362$ & $2.261$ & $0.384$ & $2.059$ & $0.293$ & $1.666$ & $0.326$& $1.549$\\
\hline
    WCMP & $0.704$& $6.698$ & $0.816$ & $5.110$ & $0.684$ & $4.512$ & $0.695$& $4.983$\\
\hline
    ECMP & $3.048$& $8.832$ & $3.086$ & $9.219$ & $3.096$ & $9.337$ & $3.085$& $9.074$\\
\hline
    \textbf{RLB-SAC-J} & $0.420$ & $2.617$ & $0.427$ & $2.484$ & $0.351$ & $2.630$ & $0.355$& $2.147$\\
\hline
    \textbf{RLB-SAC-G} & $0.382$ & $2.684$ & $0.396$ & $2.534$ & $0.340$ & $2.748$ & $0.322$& $2.136$\\
\hline
    \textbf{RLB-SAC-B} & $0.369$ & $2.791$ & $0.380$ & $2.787$ & $0.325$ & $3.230$ & $0.345$& $2.197$\\
\bottomrule
\end{tabular}
\label{tab:compare_reward_option}
\vskip -.1in
\end{table}

\section{Conclusion and Future Work}
\label{sec:conclusion}

Promising as RL techniques are, it is challenging to adapt them to realistic systems, where, in the context of network load balancing, agents have limited observation over the environment and where periodic action updates may not be able to catch up with system dynamics.
This paper takes a first step in studying the application of RL on network load balancing problems starting from the simplest assumptions that tasks are identical, towards more sophisticated (and perhaps more realistic) setups.
The preliminary results of the proposed RLB-SAC LB show promise in the proposed RL-based load balancing framework comparing with baseline LB approaches (ECMP, WCMP, LSQ, SED).
The design of the reward function requires further investigations so that RLB-SAC achieves better performance and improves generalization ability even under more realistic configurations.
In presence of multiple LB agents and multiple application servers, the scalability of RLB-SAC is limited when the action space and the number of agents grow, which makes it an interesting direction for further research on adapting different RL models and algorithms.

\bibliographystyle{unsrt}
\bibliography{reference}

\appendix

\section{Hyperparameters}
\label{app:hyperparameter}

\begin{table}[htbp]
\caption{Hyperparameters in RL-based LB.}
\label{tab:rl_params}
\resizebox{\textwidth}{!}{ 
\begin{tabular}{cccccc}
\toprule
                           & \multirow{2}{*}{Hyperparameter} & \multicolumn{4}{c}{Experiments}                                                    \\ \cline{3-6} 
                           &                                 & Traffic Rate      & TCT Distribution       & Reward Option     & Scalability       \\ \cline{2-6} 
\multirow{6}{*}{SAC}       & Learning rate                   & $1\times 10^{-3}$ & $1\times 10^{-3}$      & $1\times 10^{-3}$ & $1\times 10^{-3}$ \\ \cline{2-6} 
                           & Batch size                      & $64$              & $64$                   & $64$              & $64$              \\ \cline{2-6} 
                           & Replay Buffer Size              & $3000$            & $3000$                 & $3000$            & $3000$            \\ \cline{2-6} 
                           & Episodes                        & $20$              & $20$                   & $20$              & $40$              \\ \cline{2-6} 
                           & Step interval                   & $0.5$s            & $0.5$s                 & $0.5$s            & $0.5$s            \\ \cline{2-6} 
                           & Target entropy                  & $-|\mathcal{A}|$  & $-|\mathcal{A}|$       & $-|\mathcal{A}|$  & $-|\mathcal{A}|$  \\ \midrule
\multirow{5}{*}{LB System} & TCT Distribution                & Identical Tasks   & Exponential            & Exponential       & Exponential       \\ \cline{2-6} 
                           & Average TCT                     & $100$ms           & $\{100$ms,$ 200$ms$\}$ & $200$ms & $200$ms\\ \cline{2-6} 
                           & First episode Duration          & $60$s             & $60$s                  & $60$s             & $60$s             \\ \cline{2-6} 
                           & Incremental Duration            & $5$s              & $5$s                   & $5$s              & $5$s              \\ \cline{2-6} 
                           & Last episode Duration           & $160$s            & $160$s                 & $160$s            & $260$s            \\ \bottomrule
\end{tabular}
}
\end{table}

\end{document}